\documentclass[a4paper]{jpconf}
\usepackage{graphicx}
\usepackage{xcolor}
\usepackage{amsmath}
\usepackage{amssymb}
\begin{document}
\title{Hadronic light-by-light contribution to the muon $g-2$}

\author{Adolfo Guevara}

\address{Departamento de F\'isica, Cinvestav del IPN, Apdo. Postal 14-740, 07000 Mexico DF. Mexico}

\ead{aguevara@fis.cinvestav.mx}

\begin{abstract}
 We have computed the hadronic light-by-light (LbL) contribution to the muon anomalous magnetic moment $a_\mu$
 in the frame of Chiral Perturbation Theory with the inclusion of the lightest resonance multiplets as dynamical fields (R$\chi$T). 
 It is essential to give a more accurate prediction of this hadronic contribution due to the future projects of J-Parc and FNAL on reducing the uncertainty in this observable. 
 We, therefore, computed the pseudoscalar transition form factor and proposed the measurement of the $e^+e^-\to\mu^+\mu^-\pi^0$ cross section 
 and dimuon invariant mass spectrum to determine more accurately its parameters. Then, we evaluated the pion exchange contribution to $a_\mu$,
 obtaining $(6.66\pm0.21)\cdot10^{-10}$. By comparing the pion exchange contribution and the pion-pole approximation to the corresponding transition form factor ($\pi$TFF)
 we recalled that 
 the latter underestimates the complete $\pi$TFF by (15-20)\%. Then, we obtained the $\eta^({}'^)$ TFF, obtaining a total contribution of 
 the lightest pseudoscalar exchanges of $(10.47\pm0.54)\cdot10^{-10}$, in agreement with previous results and with smaller error.
\end{abstract}

\section{Introduction}

 Ever since the measurement of the electron magnetic moment in the splitting of the ground states of deuterium and molecular hydrogen \cite{Rabi}, the anomalous 
 magnetic moment has been an ever more stringent test of the underlying theory governing the interactions among elemental particles; giving 
 us the lead from a way of renormalizing QED \cite{Schwinger} to an outstanding confirmation of QFT with QED contributions \cite{Kinoshita} up to order $\left(\frac{\alpha}{\pi}\right)^6$. 
 In this spirit, the $a_\mu$ has been seen as a very stringent test of beyond standard model physics (BSM). With the most recent measurements \cite{BH}, a deviation 
 from the Standard Model (SM) results would imply a contribution from BSM with a scale $\sim$ 100 TeV (assuming an interaction $\sim$ 1). The current discrepancy \cite{pdg} 
 of 3.6$\sigma$ and future plans on measuring more accurately this observable force theorists to make more precise predictions of SM contributions to the $\mu$ 
 anomalous magnetic moment.\\
 
 \begin{figure}[h]\label{Had}
  \centering\includegraphics[scale=0.55]{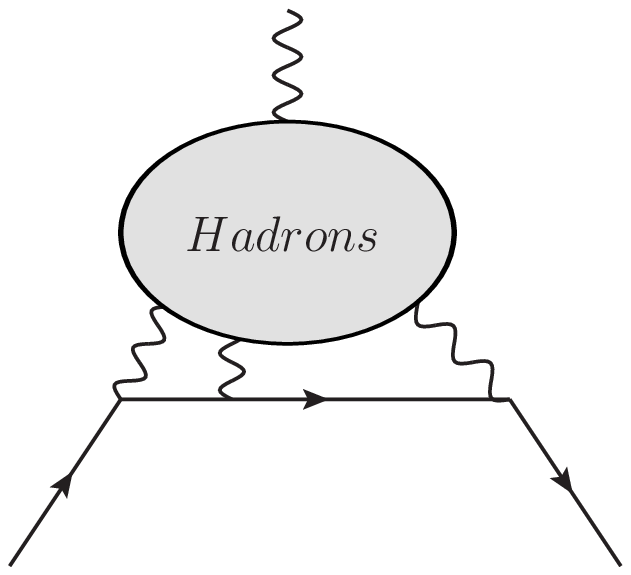}\hspace*{12ex}\includegraphics[scale=0.51]{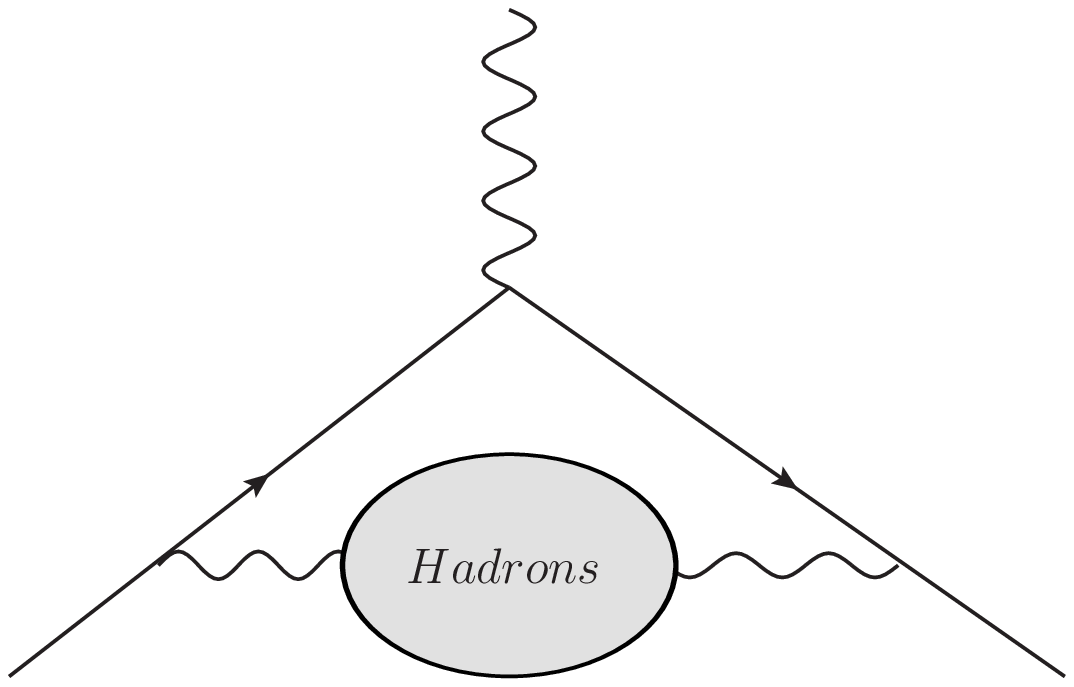}\caption{The two hadronic contribution to $a_\mu$: Hadronic 
  Light-by-Light (left) and Hadron Vacuum Polarization (right).}
 \end{figure}
 
\begin{figure}[h]
  \centering\includegraphics[scale=0.78]{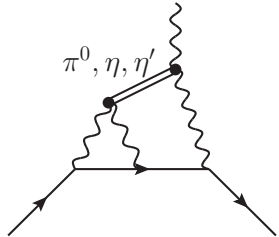}\caption{Pseudoscalar exchange contribution to the HLbL.}\label{pilbl}
 \end{figure}

 Within the SM, the contributions to the $a_\mu$ that have a greater uncertainty are the hadronic ones \cite{pdg}. This is due to the fact that the underlying theory cannot be taken 
 perturbatively in the whole energy range of the quark loop integrals, forcing theorists to compute these contributions using effective field theories (EFT) based 
 on symmetries of Quantum Chromodynamics (QCD). This hadronic contribution can be splitted into two sub-contributions, the Hadron Vacuum Polarization (HVP) and the 
 Hadronic Light-by-Light (HLbL), shown in Fig. \ref{Had}. We analize the latter one by studying the $P\gamma^*\gamma^*$ interaction through its form factor (also called P 
 transition form factor, PTFF), which gives the leading contribution to the HLbL through a pseudoscalar exchange diagram shown in Fig. \ref{pilbl}. At low energies (\textit{i.e.} in the chiral limit), 
 the prediction for the $\pi$TFF has been confirmed by the measured rate of $\pi^0\to\gamma\gamma$ decays \cite{pdg}. On the other hand, the prediction for a nearly on-shell 
 photon and one with very large virtuality seems to be at odds with measurements at B-factories \cite{BaBar, Belle}. These two limits have ruled the way of constructing 
 the form factor to describe interactions in the intermediate energy region, where hadronic degrees of freedom play a crucial role. The EFT we use to compute the TFF is 
 Resonance Chiral Theory (R$\chi$T) \cite{EG, KN}, which makes use of short-distance QCD predictions to obtain the parameters of the theory in terms of known 
 constants. In this work, we fit one of the parameters in the $\pi$TFF with the B-factories data and, using this information, we predict the 
 $\eta^({}'^)$TFF and then obtain the contribution to the $a_\mu$ using these form factors.
 
\section{Theoretical Framework}
 Chiral Perturbation Theory ($\chi$PT) \cite{ChPT} is the EFT dual to QCD at low energies \cite{CCWZ}. It is based on an expansion in powers of momenta and 
 masses of the lightest pseudoscalar mesons over the chiral 
 symmetry breaking scale ($\Lambda\sim$ 1 GeV). Thus, the theory fails to be reliable at energies around 1 GeV; furthermore, when other mesons become relevant degrees of 
 freedom ($\Lambda\gtrsim M_\rho$) the theory is no longer applicable. A generalization of $\chi$PT is obtained using $1/N_C$ as an expansion parameter \cite{NC} to include 
 resonances as dynamical degrees of freedom. The theory that incorporates these elements is Resonance Chiral Theory (R$\chi$T) \cite{EG, KN}, which requires unitary 
 symmetry for the resonance multiplets. No \textit{a priori} assumptions are made with respect to the role of resonances in this theory, therefore one obtains naturally 
 Vector Meson Dominance \cite{VMD} as a dynamical result of the theory \cite{EG}. The final ingredient of the theory comes from QCD behavior at short distances, which 
 constraints a great amount of free parameters in the theory. 
 
\section{The $\pi\gamma^*\gamma^*$ form factor in R$\chi$T} 
 In this framework, the form factor we obtain \cite{us} is

 \begin{multline}\label{FullTFF}
 \mathcal{F}_{\pi^0\gamma^*\gamma^*}(p^2,q^2,r^2)=\frac{2r^2}{3F}\left[-\frac{N_C}{8\pi^2r^2}+4F_V^2\frac{d_3(p^2+q^2)}{(M_V^2-p^2)(M_V^2-q^2)r^2}
  +\frac{4F_V^2d_{123}}{(M_V^2-p^2)(M_V^2-q^2)}\right.\\+\frac{16F_V^2P_3}{(M_V^2-p^2)(M_V^2-q^2)(M_P^2-r^2)}
  \left.-\frac{2\sqrt{2}}{M_V^2-p^2}\left(\frac{F_V}{M_V}\frac{r^2c_{1235}-p^2c_{1256}+q^2c_{125}}{r^2}+\frac{8P_2F_V}{(M_P^2-r^2)}
  \right)\right.\\\left.+(q^2\leftrightarrow p^2)\right].
 \end{multline}
 
 It contains contributions from the pseudoscalar resonances (as can be seen through the couplings $P_2$ and $P_3$) which need to be taken into account 
 to obtain consistent short distance constraints \cite{KN, RSC}. All the parameters, but one, can be obtained through these constraints. $P_3$ cannot be obtained requiring 
 high energy constraints, therefore it is fitted using the combined analyses of $\pi(1300)\to\gamma\gamma$ and $\pi(1300)\to\rho\gamma$ decays as given in 
 references \cite{KN, us}. The consistent short distance constraints on the resonance couplings in the odd-intrinsic parity sector can be seen in refs \cite{us, RSC}. Thus, 
 we obtain 
 \begin{equation}
  P_3 = (-1.2\pm0.3)\cdot10^{-2}\text{ GeV}^2.
 \end{equation}
 \begin{figure}[h]
  \centering\includegraphics[scale=0.34,angle=-90]{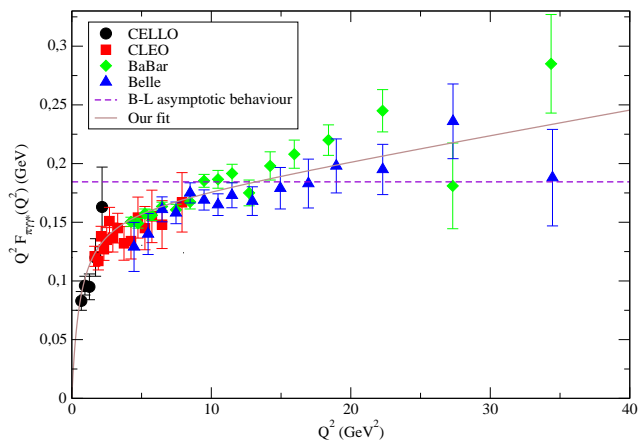}\caption{Our best fit compared to CELLO, CLEO, BaBar and Belle data for the $\pi$TFF.}\label{piTFF}
 \end{figure}

 On the other hand, the $\pi$TFF does not fit very well experimental data \cite{BaBar, Belle} when $P_2$ is constrained by the short distance prediction. Therefore we 
 allowed for it a small variation in a fit to Babar and Belle data of this form factor, where they measure the $\pi$TFF spectrum in a kinematical configuration that ensures 
 that one of the photons is on-shell and the other is virtual. The form factor for such a configuration is given by taking~\footnote{By the kinematical configuration in which 
 the process is chosen to be measured, the momenta of both photons are space-like.}
  $p^2\to0$ and $Q^2=-q^2$ in eq. (\ref{FullTFF})
 \begin{equation}\label{TFFQ2}
  \mathcal{F}_{\pi^0\gamma^*\gamma}(Q^2)=-\frac{F}{3}\frac{Q^2(1+32\sqrt{2}\frac{P_2F_V}{F^2})+\frac{N_C}{4\pi^2}\frac{M_V^4}{F^2}}{M_V^2(M_V^2+Q^2)}.
 \end{equation}
 We keep a very conservative 10\% uncertainty from the asymptotic value of $F_V$ around its predicted value~\cite{RSC} of $\sqrt{3}F$.
 Fig. \ref{piTFF} shows our best fit, with which we obtain 
 \begin{equation}
  P_2=(-1.13\pm0.12)\cdot10^{-3}\text{ GeV,\hspace*{15ex}}\chi^2/dof=1.01\,.
 \end{equation}
\begin{figure}[h]
 \centering\includegraphics[scale=0.33,angle=-90]{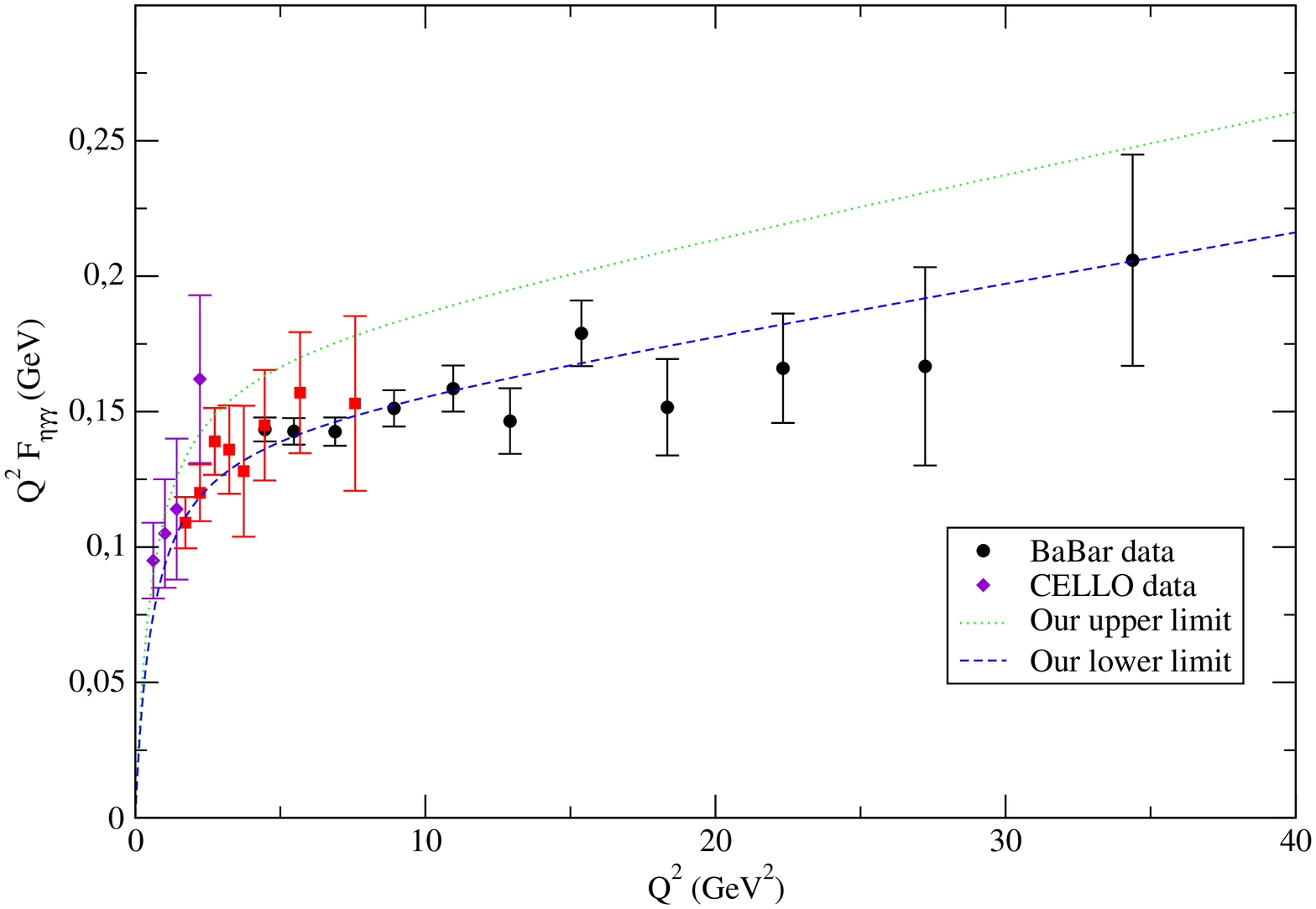}\includegraphics[scale=0.33,angle=-90]{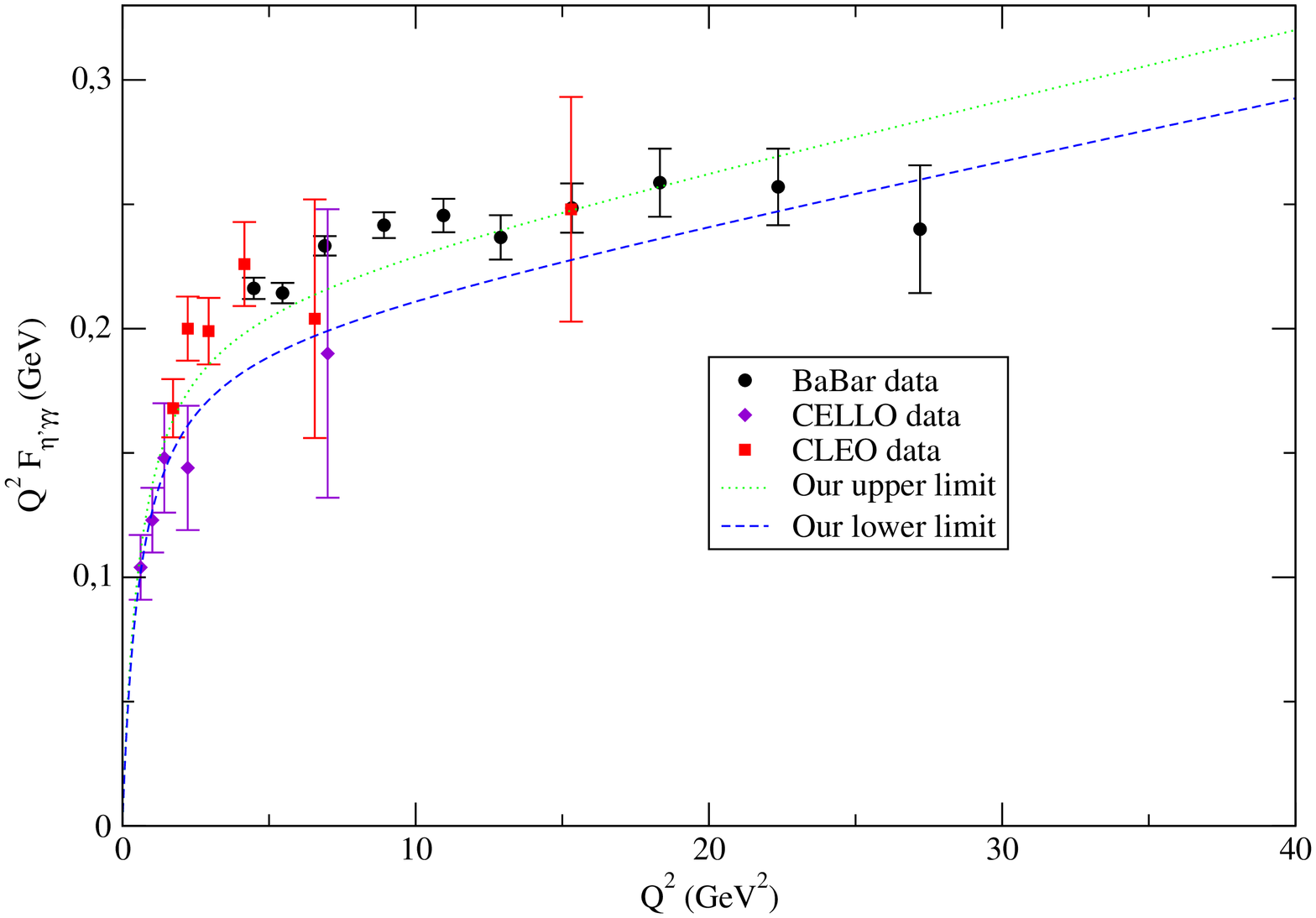}\caption{Our prediction of the $\eta$TFF (left) 
 and $\eta'$TFF (right) compared to CELLO, CLEO and BaBar data and using the parameters found with the $\pi$TFF.}\label{etaTFF}
\end{figure}

\section{The pseudoscalar exchange contribution to the $a_\mu^{HLbL}$}

 Once all the parameters in the $\pi$TFF are determined, we insert the full off-shell TFF in the relations given in \cite{Knecht}
 obtaining thus 
 \begin{equation}
  a_\mu^{\pi^0LbL}=(5.75\pm0.06)\cdot10^{-10}\text{ on-shell }\pi^0\hspace*{8.7ex}a_\mu^{\pi^0LbL}=(6.66\pm0.21)\cdot10^{-10} \text{ whole }\pi^0\text{TFF}.
 \end{equation}
 
 \begin{table}[h]
 \centering\caption{Our result compared with other results obtained through different methods.}
  \begin{tabular}{cc}\br
$a_\mu^{\pi^0 LbL}\cdot{10}$& Model and Reference \\\mr
 \small{$5.58\pm0.05$} & \small{Extended Nambu-Jona-Lasinio \cite{Bijnens}}\\
\small{$5.56\pm0.01$}&\small{Naive VMD \cite{Hayakawa}}\\
\small{$5.8\pm1.0$} &\small{Large $N_C$ with two vector multiplets $\pi$-pole \cite{Knecht}}\\
\small{$7.2\pm1.2$} &\small{$\pi$-exchange contribution \cite{Jegerlehner}}\\
\small{$6.54\pm0.25$}&\small{Holographic models of QCD \cite{Cappiello}}\\
\small{$6.58\pm0.12$}&\small{Lightest pseudoscalar and vector resonance saturation \cite{KN}}\\
\small{$6.49\pm0.56$}&\small{Rational approximants \cite{Masjuan}}\\
\small{$5.0\pm0.4$}&\small{Non-local chiral quark model \cite{Dorokhov}}\\
\mr
\small{$5.75\pm0.06$}&\small{Our result with on-shell $\pi^0$ \cite{us}}\\
 $6.66\pm0.21$& Our result whole $\pi^0$TFF \cite{us} \\\br
   \end{tabular}\label{comp}
  \end{table}

 This clearly shows that assuming an on-shell pion in the $a_\mu^{HLbL}$ underestimates the contribution in $\sim15\%$, and the error by a factor of 4. The
 uncertainty comes mainly from the error in $F_V$, $P_3$ and in a chiral correction from very-low energy physics. We compare our result with previous 
 results in table \ref{comp}. The form factor for the $\eta$ and $\eta'$ can be obtained with the $\pi$TFF through eq. (\ref{eta-etap}) with the minus sign 
 for the case of the $\eta$.
 \begin{equation}\label{eta-etap}
  \mathcal{F}_{\eta^({}'^)\gamma^*\gamma^*}=\left(\frac{5}{3}C_{q}^({}'{}^)\mp\frac{\sqrt{2}}{3}C_{s}^({}'{}^)\right)\mathcal{F}_{\pi^0\gamma^*\gamma^*}
 \end{equation}

  With this, we can compute the whole pseudoscalar exchange contribution to the $a_\mu^{HLbL}$, shown in table \ref{amu} which includes other sub-leading HLbL 
  contributions.
  
  \begin{table}[h]
  \centering\caption{Comparison of our contributions of the full $a_\mu^{HLbL}$ to previous determinations.}
   \begin{tabular}{cc}\br
    $a_\mu^{HLbL}\cdot10^{10}$ & Contributions\\\mr
    $11.6\pm4.0$&F. Jegerlehner and A. Nyffeler \cite{Jegerlehner}\\
    $10.5\pm2.6$&Prades, De Rafael and Vainshtein\footnote{
    Prades {\it et al.} only includes the {\it charm} loop in the heavy quark loop evaluation.} \cite{PdRV}\\
    $11.8\pm2.0$&Our contribution  \cite{us}\\\br
   \end{tabular}\label{amu}
 \end{table}
  \section{Genuine probe of $\pi$TFF}

\begin{figure}[h]
  \centering\includegraphics[scale=0.32,angle=-90]{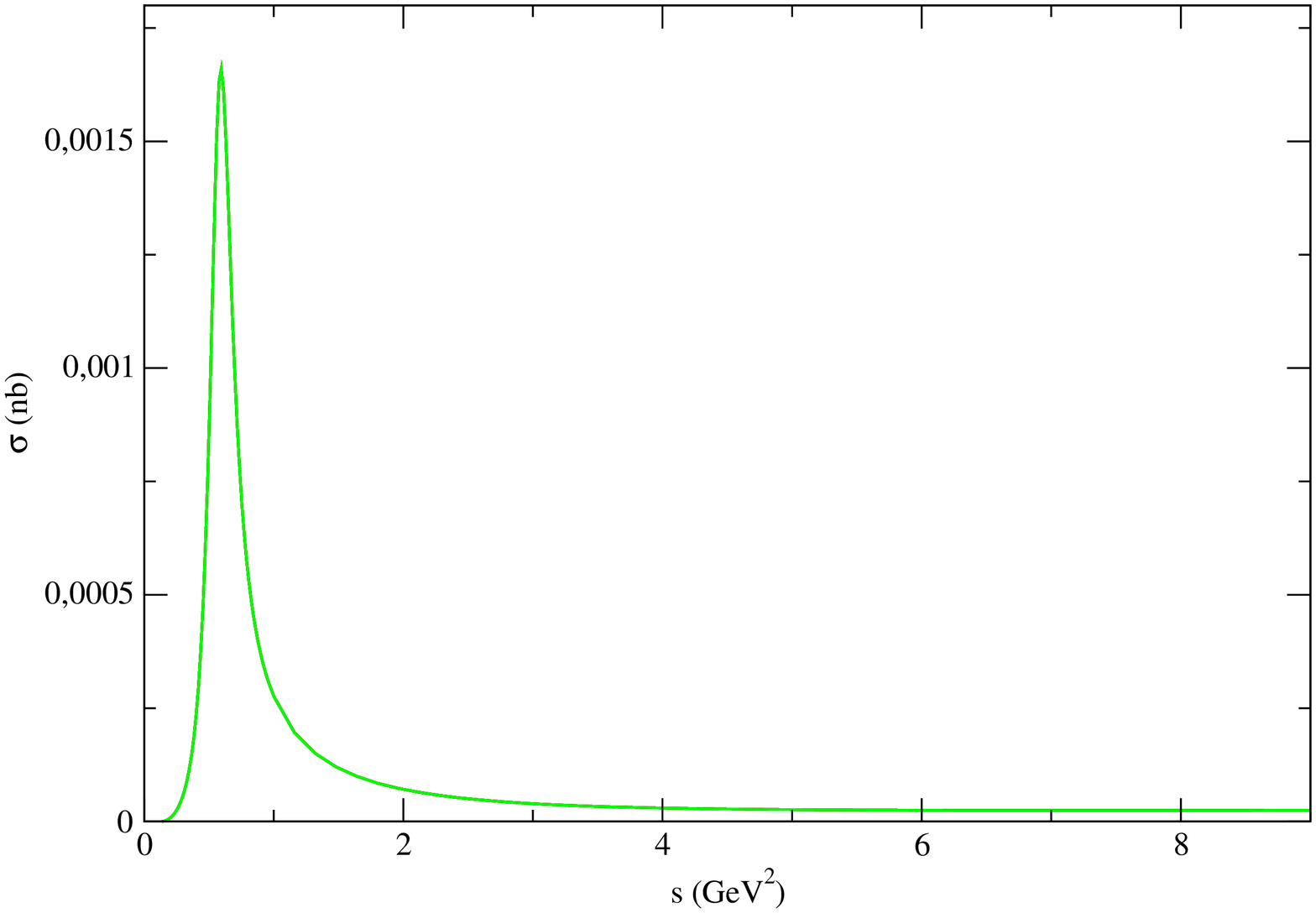}\hspace*{1ex}\includegraphics[scale=0.32,angle=-90]{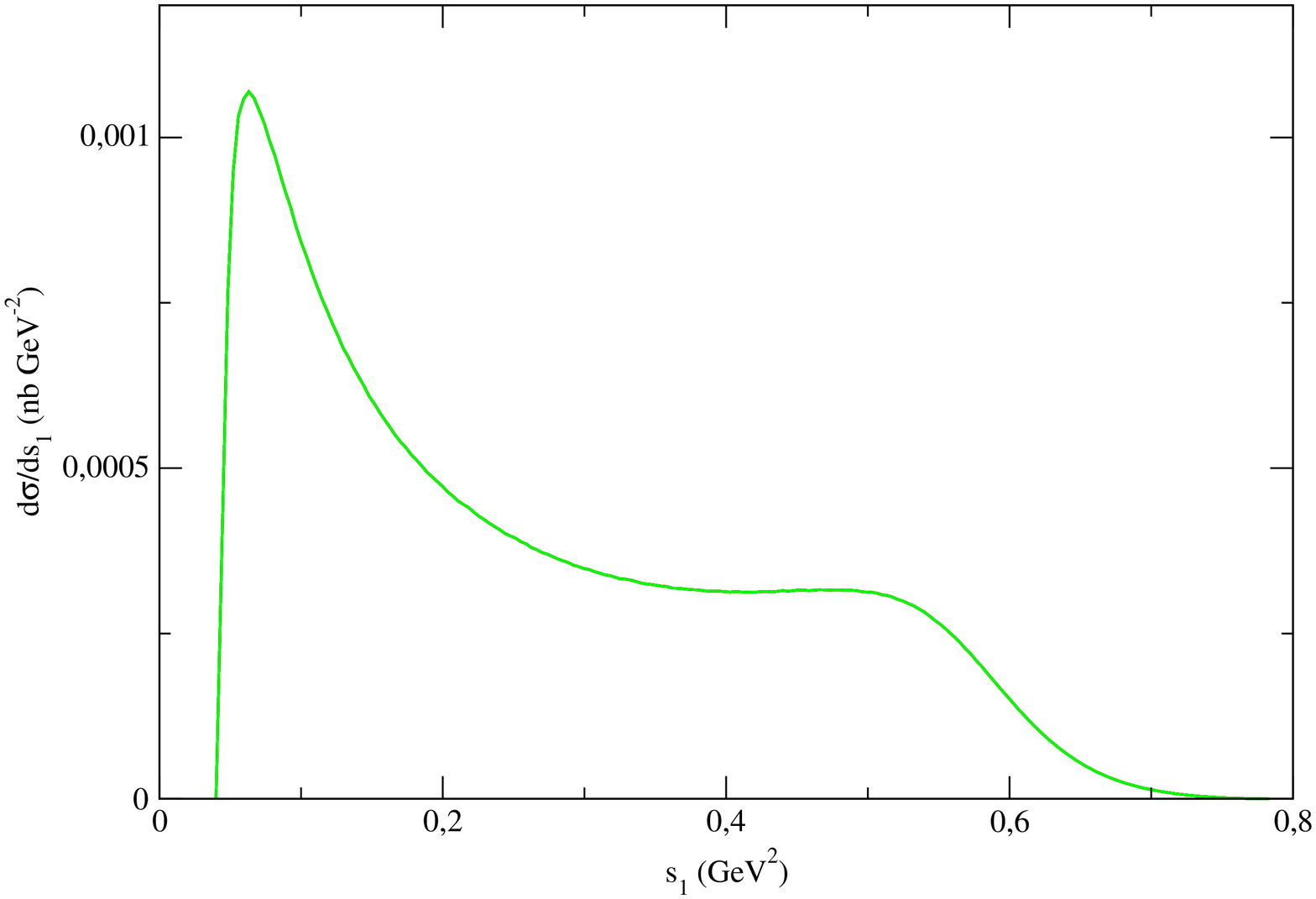}\caption{Our prediction for $\sigma(s)$ (left)
  and for $\frac{s\sigma}{ds_1} \text{ with } s=1.02\text{ GeV}^2$, the error bands cannot be appreciated in these plots.}\label{NO}
 \end{figure}

 All the experimental observables available to fit the parameters in the $\pi$TFF so far need an on shell photon and/or have photons with space-like momenta, 
 while the HLbL contribution to the $a_\mu$ has both photons with time-like momenta. The photons in the process we study in this section, namely 
 $\sigma(e^+e^-\to\gamma^*\to\pi^0\gamma^*\to\mu^+\mu^-\pi^0)$, both have time-like momenta and can be measured at very high photon virtualities by KLOE collab. for 
 $q^2\sim1$ GeV$^2$ and Belle-II collab. for $q^2\sim10.5\text{ GeV}^2$. With the $\pi$TFF parameters fully determined, the prediction we obtain is shown in Fig. \ref{NO}
 \section{Conclusion}
 We found the pseudoscalar exchange contribution to the $a_\mu^{HLbL}$ with a very competitive uncertainty and consistent with other theoretical models; improving 
 the analysis by including high energy constraints not realized in the reference \cite{KN} and also using Belle data released after the reference was published.
 Our error estimate is also more robust, since in addition to the errors of the resonance couplings, we have also included the uncertainty due to the value of the 
 $\pi$TFF at very low energies.\\
 
 We also obtained the first prediction for the cross section $\sigma(e^+e^-\to\mu^+\mu^-\pi^0)$, which might be measured in KLOE-2 and Belle-II. The measurement of this 
 observable would be an interesting way of trying to reduce the error in the parameters of the $\pi$TFF. This, may also help to reduce the uncertainty on the mixing 
 parameters between the $\eta$ and $\eta'$ mesons.
 
 \ack{The author wishes to thank the DPyC and Cinvestav for financial support.}

\end{document}